\documentclass{article}
\usepackage{fleqn,espcrc2}

\usepackage{graphicx}
\def\eg{{e.g., }}
\def\ie{{i.e., }}
\def\etal{{et al., }}
\def\et{{et al. }}
\def\etc{{etc.}}

\def\'{^{\prime}}

\def\hq{{\hat q}}

\def\hmpc{{\, {\rm h}^{-1}~\rm Mpc}}

\def\kms{{\rm~km~s^{-1}}}

\def\kpc{{\rm~kpc}}
\def\mpc{{\rm~Mpc}}

\def\spose#1{\hbox to 0pt{#1\hss}}
\def\lta{\mathrel{\spose{\lower 3pt\hbox{$\mathchar"218$}}
     \raise 2.0pt\hbox{$\mathchar"13C$}}}
\def\gta{\mathrel{\spose{\lower 3pt\hbox{$\mathchar"218$}}
     \raise 2.0pt\hbox{$\mathchar"13E$}}}
\def\ge{\mathrel{\spose{\lower 3pt\hbox{$-$}}
     \raise 2.0pt\hbox{$\mathchar"13E$}}}
\def\le{\mathrel{\spose{\lower 3pt\hbox{$-$}}
     \raise 2.0pt\hbox{$\mathchar"13C$}}}
\newcommand{\AmS}{{\protect\the\textfont2
  A\kern-.1667em\lower.5ex\hbox{M}\kern-.125emS}}

\title{The Cosmic Background Radiation {\it circa} $\nu$2K}

\author{J. Richard Bond, Dmitry Pogosyan, Simon Prunet \address{CIAR Cosmology Program, Canadian Institute
for Theoretical Astrophysics, \\
        60 St. George St., Toronto, ON M5S 3H8, Canada}
        and
        the MaxiBoom Collaboration\address{See Jaffe \et 2000 \cite{jaffe00} for the full author and institution
        list.\\
        {\tt \small CITA-2000-63, in Proc. Neutrino 2000 (Elsevier), eds.
J. Law, J. Simpson\normalsize}}}

\begin{document}

\begin{abstract}
We describe the implications of cosmic microwave background (CMB)
observations and galaxy and cluster surveys of large scale
structure (LSS) for theories of cosmic structure formation,
especially emphasizing the recent Boomerang and Maxima CMB balloon
experiments.  The inflation-based cosmic structure formation
paradigm we have been operating with for two decades has never
been in better shape. Here we primarily focus on a simplified
inflation parameter set, $\{\omega_b,\omega_{cdm},\Omega_{tot},
\Omega_\Lambda,n_s,\tau_C, \sigma_8\}$. Combining all of the
current CMB+LSS data points to the remarkable conclusion that the
local Hubble patch we can access has little mean curvature
($\Omega_{tot} = 1.08\pm 0.06$) and the initial fluctuations were
nearly scale invariant ($n_s =1.03 \pm 0.08$), both predictions of
(non-baroque) inflation theory. The baryon density is found to be
slightly larger than that preferred by independent Big Bang
Nucleosynthesis estimates ($\omega_b\equiv \Omega_b {\rm h}^2 =
0.030\pm 0.005$ cf. $0.019\pm 0.002$). The CDM density is in the
expected range ($\omega_{cdm}=0.17 \pm 0.02$). Even stranger is
the CMB+LSS evidence that the density of the universe is dominated
by unclustered energy akin to the cosmological constant
($\Omega_\Lambda = 0.66 \pm 0.06$), at the same level as that
inferred from high redshift supernova observations. We also sketch
the CMB+LSS implications for massive neutrinos. \vspace{1pc}
\end{abstract}

\maketitle

\section{CMB Anisotropies \& Distortions}

\subsection{What Was Revealed in 1999/2000}

\begin{figure}
%[htb]
\hspace{-0.8truecm}
\includegraphics[scale=0.50]{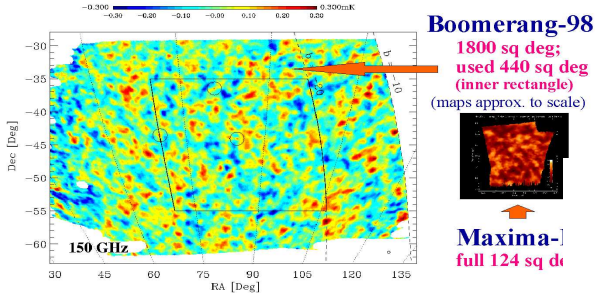} \vskip
-1.0 truecm
 \caption{\small The Boomerang 150A GHz bolometer map (one out of 16)
is shown. Of the entire 1800 square degrees covered (not all of
which is shown), only the interior 440 sq. degs. within the
central rectangle were used in our analysis, \ie in all less than
5\% of the data.  The 124 square degree Maxima-I map, drawn to
scale, is also shown. Bandpowers derived from these maps and their
noise correlation matrices (Figure~\ref{fig:CLdat}) were used to
derive the parameters given in Table~\ref{tab:exptparams}.
\normalsize } \label{fig:map}
%  \end{center}
\end{figure}

\begin{figure}
%[htb]
\hspace{-1.2truecm}
\includegraphics[scale=0.65]{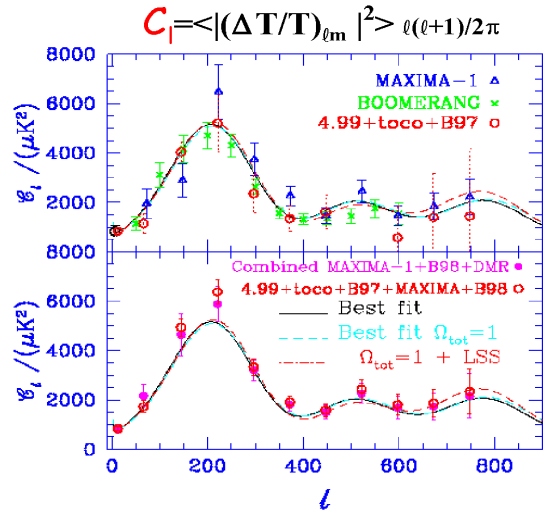}
\vskip -1.0 truecm \caption{\small The ${\cal C}_\ell$ (defined in
terms of CMB temperature anisotropy multipoles, $(\Delta
T/T)_{\ell m}$, as indicated) grouped in bandpowers for prior-CMB
experiments, including TOCO and the North American Boomerang test
flight, (squares)  are contrasted with the Boomerang-98 LDB
(crosses) and Maxima-I (triangles) results in the upper panel. The
lower panel shows the optimally combined power spectra of
Boomerang+Maxima+DMR (squares) and contrasts it with that for
Boomerang+Maxima+prior-CMB (circles), showing that including the
prior experiments does not make a large difference to the results.
Best-fit models for arbitrary $\Omega_{tot}$ and for the
restriction to the flat case are shown in both panels. \normalsize
} \label{fig:CLdat}
%  \end{center}
\end{figure}

As we penetrate the CMB, we go from a nearly perfect blackbody of
$2.725 \pm 0.002\, K$ \cite{matherTcmb}, to a $3.372 \pm 0.007\,
mK$ dipole pattern associated with the 300 $\kms$ flow of the
earth in the CMB, to the now familiar motley pattern of
anisotropies associated with $2 \le \ell \lta 20$ multipoles at
the $30 \mu K$ level revealed by COBE at $7^\circ$ resolution,
then at higher $\ell$ in some 19 other experiments  (though most
with many fewer resolution elements than the 600 or so for COBE).
The picture was dramatically improved this year, as results were
announced first in summer 99 from the ground-based TOCO experiment
in Chile \cite{toco98}, then in November 1999 from the North
American balloon test flight of Boomerang \cite{mauskopf99}, then
in April 2000 from the first CMB long duration balloon (LDB)
flight, of Boomerang \cite{debernardis00}, followed in May 2000
from the night flight of Maxima \cite{MAXIMA1}. Boomerang's best
resolution was $10^\prime$, about 40 times better than that of
COBE, with tens of thousands of resolution elements. Maxima had a
similar resolution but covered an order of magnitude less sky.
Both experiments were designed to reveal the {\it primary}
anisotropies of the CMB, those which can be calculated using
linear perturbation theory. What we see in Fig.~\ref{fig:map} are
two images of soundwave patterns that existed about 300,000 years
after the Big Bang, when the photons were freed from the plasma.
The visually evident structure on degree scales is even more
apparent in the power spectra of the Fourier transform of the maps
(Fig.~\ref{fig:CLdat}), which show a dominant (first acoustic)
peak, a less prominent (or non-existent) second one, and the hint
of a third one from Maxima. In the following, we describe the
implications of these results for cosmology. Space constraints
preclude adequate referencing here, but these are given in
\cite{bh95,lange00,jaffe00}.

We are only at the beginning of the high precision CMB era.
HEMT-based interferometers are already in place taking data: the
VSA (Very Small Array) in Tenerife, the CBI (Cosmic Background
Imager) in Chile, DASI (Degree Angular Scale Interferometer) at
the South Pole, where the bolometer-based single dish ACBAR
experiment will operate this year. Other LDBs will be flying
within the next few years: Arkeops, Tophat, Beast/Boost; and in
2001, Boomerang will fly again, this time concentrating on
polarization. As well, MAXIMA will fly as the
polarization-targeting MAXIPOL. In April 2001, NASA will launch
the all-sky HEMT-based MAP satellite, with $12^\prime$ resolution.
Further downstream in 2007, ESA will launch the
bolometer+HEMT-based Planck satellite, with $5^\prime$ resolution.

\subsection{Primary \& Secondary Processes}

As we peer back in time, we reach a fuzzy wall at redshift $z \sim
1100$, the photon decoupling "surface", through which the Universe
passed from optically thick to thin to Thomson scattering over a
physical distance $\sim 10 \kpc$, \ie $\sim 10 \mpc$ comoving.
Prior to this, acoustic wave patterns in the tightly-coupled
photon-baryon fluid on scales below the "sound crossing distance"
at decoupling, $\lta 100 \kpc$ physical, $\lta 100 \mpc$ comoving,
were viscously damped, strongly so on scales below the $\sim 10
\kpc$ thickness over which decoupling occurred. After, photons
freely-streamed along geodesics to us, mapping (through the
angular diameter distance relation) the post-decoupling spatial
structures in the temperature to the angular patterns we observe
now as the {\it primary} CMB anisotropies. The maps are images
projected through the fuzzy decoupling surface of the acoustic
waves (photon bunching), the electron flow (Doppler effect) and
the gravitational potential peaks and troughs ("naive" Sachs-Wolfe
effect) back then. Free-streaming along our (linearly perturbed)
past light cone leaves the pattern largely unaffected, except that
temporal evolution in the gravitational potential wells as the
photons propagate through them leaves a further $\Delta T$
imprint, called the integrated Sachs-Wolfe effect. Intense
theoretical work over three decades has put accurate calculations
of this linear cosmological radiative transfer on a firm footing,
and there is a speedy, publicly available and widely used code for
evaluation of anisotropies in a wide variety of cosmological
scenarios, ``CMBfast'' \cite{cmbfast}.

{\it Secondary anisotropies} associated with nonlinear effects
also leave an imprint. The photons were weakly-lensed by
intervening mass, were Thompson-scattered by the flowing gas once
it became "reionized" at $z \sim 20$, and were occasionally
Compton-upscattered, with mean fractional  energy shifts $
4T_{gas}/m_ec^2$, by nonlinear hot gas (thermal Sunyaev-Zeldovich
or Compton cooling effect); the flowing cluster induces a further
kinetic SZ effect. Of these secondary processes, all have the same
spectral signature of a perturbed blackbody, with a
$\lambda$-independent thermodynamic temperature perturbation,
$\Delta T(\lambda, \hq)$ --- except for the thermal SZ effect.

The SZ  spectral distortion is related to the Compton $y$
parameter, which is proportional to the line of sight integral of
the electron pressure: $\Delta T$ is negative, $-2y$,  on the
Rayleigh-Jeans side and positive on the Wien side,
$(E_\gamma/T_{cmb})y$, passing through zero at $1378 \mu {\rm m}$
or 218 GHz. The COBE FIRAS experiment limits the energy injection
that leads to the thermal SZ pattern to be ${\delta E_{Compton\
cool} / E_{cmb}} = 4y < 6.0 \times 10^{-5}$ (95\% CL)
\cite{matherTcmb}. It is routine now to see the (concentrated)
fluctuations in $y$ from clusters, though the sky-average $y$ is
predicted to be $\lta 10^{-5}$. SZ anisotropies have been probed
by single dishes, the OVRO and BIMA mm arrays, and the Ryle
interferometer. A number of HEMT-based interferometers being built
are more ambitious: AMI (Britain), the JCA (Chicago), AMIBA
(Taiwan), MINT (Princeton). As well, other kinds of
bolometer-based experiments will be used to probe the SZ effect,
including the CSO (Caltech submm observatory) with BOLOCAM on
Mauna Kea, ACBAR at the South Pole, the  LMT (large mm telescope)
in Mexico, and the LDB BLAST.

A secondary anisotropy not associated with the CMB photons is
stellar and accretion disk radiation reprocessed by dust into the
infrared and redshifted into the submm, leading to a Wien tail
distortion of the CMB blackbody. This has been found in the COBE
FIRAS and shorter wavelength DIRBE data, with energy at a level
about twice that in optical light. Anisotropies from dust emission
from high redshift galaxies are being targeted by the JCMT with
the SCUBA bolometer array, the CSO (soon with BOLOCAM), the OVRO
mm interferometer, the SMA (submm array) on Mauna Kea, the LMT,
the ambitious US/ESO ALMA array in Chile, the LDB BLAST, and ESA's
FIRST satellite. About $50\%$ of the submm background has so far
been identified with sources that SCUBA has found.

 ${\cal C}_\ell$'s from nonlinearities in the gas at high redshift
are concentrated at high $\ell$, but for most viable models are
expected to be a small contaminant. Similarly, Thomson scattering
from gas in moving clusters also has a small effect on ${\cal
C}_\ell$ (although it should be measurable in individual clusters,
\eg  with BOLOCAM and with Planck). The effect of lensing is to
smooth slightly the acoustic peaks and troughs of
Fig.~\ref{fig:CLdat} and induce small scale non-Gaussian effects.

\subsection{Boomerang \& Maxima}

Boomerang carried a  1.2m telescope with 16 bolometers cooled to
0.3K in the focal plane aloft from McMurdo Bay in Antarctica in
late December 1998, circled the Pole for 10.6 days and landed just
50 km from the launch site, only slightly damaged. In
\cite{debernardis00}, maps at 90, 150 and 220 GHz showed the same
spatial features and the intensities were shown to fall precisely
on the CMB blackbody curve. The fourth frequency channel at 400
GHz is dust-dominated. Fig.~\ref{fig:map} shows the 150 GHz map
derived using only one of the 16 bolometers. Although Boomerang
altogether probed 1800 square degrees, only the region in the
rectangle covering 440 square degrees was used in the analysis
described in \cite{lange00,jaffe00} and this paper.
Fig.~\ref{fig:map} also shows  the 124 square degree region of the
sky (in the Northern Hemisphere) that Maxima-1 probed. Though
Maxima was not an LDB, it did so well because its bolometers were
cooled even more than Boomerang's,  leading to higher sensitivity
per unit observing time; further, all frequency channels were used
in creating its map.

Analyzing Boomerang and other experiments involves a pipeline that
takes (1) the timestream in each of the bolometer channels coming
from the balloon plus information on where it is pointing and
turns it into (2) spatial maps for each frequency characterized by
average temperature fluctuation  values in each pixel
(Fig.~\ref{fig:map}) and a pixel-pixel correlation matrix
characterizing the noise, from which various statistical
quantities are derived, in particular (3) the temperature power
spectrum as a function of multipole (Fig.~\ref{fig:CLdat}),
grouped into bands, and two band-band error matrices which
together determine the full likelihood distribution of the
bandpowers \cite{bjk9800}. Fundamental to the first step is the
extraction of the sky signal from the noise, using the only
information we have, the pointing matrix mapping a bit in time
onto a pixel position on the sky.

There is generally another step in between (2) and (3), namely
separating the multifrequency spatial maps into the physical
components on the sky: the primary CMB, the thermal and kinematic
Sunyaev-Zeldovich effects, the dust, synchrotron and
bremsstrahlung Galactic signals, the extragalactic radio and
submillimetre sources. The strong agreement among the Boomerang
maps indicates that to first order we can ignore this step, but it
has to be taken into account as the precision increases. The
Fig.~\ref{fig:CLdat} map  is consistent with a Gaussian
distribution, thus fully characterized by just the power spectrum.
Higher order (concentration) statistics (3,4-point functions,
\etc) tell us of non-Gaussian aspects, necessarily expected from
the Galactic foreground and extragalactic source signals, but
possible even in the early Universe fluctuations. For example,
though non-Gaussianity occurs only in the more baroque inflation
models of quantum noise, it is a necessary outcome of
defect-driven models of structure formation. (Peaks compatible
with Fig.~\ref{fig:CLdat} do not appear in non-baroque defect
models, which now appear unlikely.)

\begin{figure}
%[htb]
\hspace{-1.2truecm}
\includegraphics[scale=0.50]{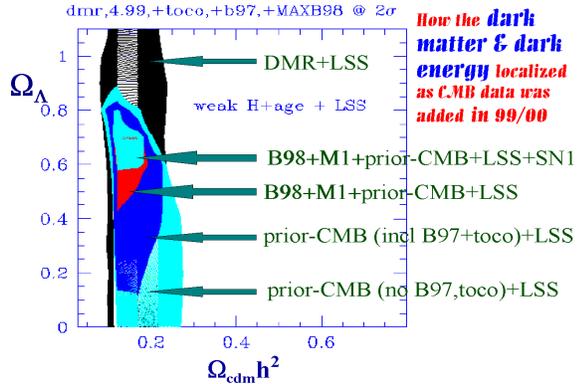}
\vskip -1.0 truecm \caption{\small 2-sigma likelihood contours for
the dark matter density and dark energy density (assuming
$w_Q=-1$, \ie a cosmological constant) are shown when
progressively more CMB data are added to the LSS  data. The
smallest 2-sigma region shows what happens when SNIa data is added
to Boomerang+Maxima+prior-CMB+LSS. In all cases, the weak-H+age
prior probability is adopted. Other variables have been
marginalized, as in Table~\ref{tab:exptparams}.  \normalsize }
\label{fig:dmde}
%  \end{center}
\end{figure}

\subsection{Parameters of Structure Formation}
\label{seccosmicparam}

 For this paper, we adopt the restricted set of 7 cosmological parameters used in
\cite{lange00,jaffe00}. Among these, we have 2 characterizing the
early universe primordial power spectrum of gravitational
potential fluctuations $\Phi$, one giving the overall power
spectrum  amplitude ${\cal P}_{\Phi}(k_n)$, and one defining the
shape, a spectral tilt $n_s (k_n) \equiv 1+d\ln {\cal P}_{\Phi}/d
\ln k$, at some (comoving) normalization  wavenumber $k_n$. We
really need another 2,  ${\cal P}_{GW}(k_n)$ and $n_t(k_n)$,
associated with the gravitational wave component. In inflation,
the amplitude ratio is related to $n_t$ to lowest order, with
${\cal O}(n_s-n_t)$ corrections  at higher order, \eg \cite{bh95}.
There are also useful limiting cases for the $n_s-n_t$  relation.
However, as one allows the baroqueness of the inflation models to
increase, one can entertain essentially any power spectrum (fully
$k$-dependent $n_s(k)$ and $n_t(k)$) if one is artful enough in
designing inflaton potential surfaces. As well, one can have more
types of modes present, \eg scalar isocurvature modes (${\cal
P}_{is}(k_n),n_{is}(k)$) in addition to, or in place of, the
scalar curvature modes (${\cal P}_{\Phi}(k_n),n_{s}(k)$). However,
our philosophy is consider minimal models first, then see how
progressive relaxation of the constraints on the inflation models,
at the expense of increasing  baroqueness,  causes the parameter
errors to open up. For example, with COBE-DMR and Boomerang, we
can probe the GW contribution, but the data are not powerful
enough to determine much. Planck can in principle probe the
gravity wave contribution reasonably well.

\begin{figure}
%[htb]
\hspace{-0.9truecm}
\includegraphics[scale=0.50]{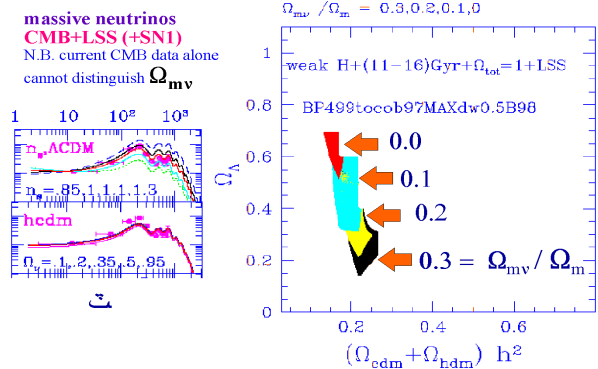}
\vskip -1.0 truecm \caption{\small 2-sigma likelihood contours in
the $(\omega_{cdm}+\omega_{hdm})$---$\Omega_\Lambda$ plane  are
shown as the fraction of matter in massive neutrinos is increased
from 0 up to 0.3, for Boomerang+Maxima+prior-CMB+LSS when the
weak-H+age + $\Omega_{tot}=1$ prior probability is adopted. The
age restriction used here, $11<$ age $<16$ Gyr,  is slightly
stronger than the $> 10$ Gyr weak-age prior adopted for the
Table~\ref{tab:exptparams} limits. The CMB data alone cannot now
discriminate among the cases --- the left plot contrasts the large
${\cal C}_\ell$ variation as $n_s$ is varied over the tiny one as
$\Omega_{hdm}$ is varied from 0 to 0.95. When LSS is added, the
contours shift in the expected direction, with more (cold+hot)
dark matter preferred, at the expense of less dark energy.
Nonetheless, significant dark energy is apparently required. Note
that the simplest interpretation of the superKamiokande data on
atmospheric $\nu_\mu$ is that $\Omega_{\nu_\tau} \sim 0.001$,
about the energy density of stars in the universe, which implies a
cosmologically negligible effect. Degeneracy between \eg $\nu_\mu$
and $\nu_\tau$ could lead to the cosmologically very interesting
$\Omega_{m\nu}\equiv \Omega_{hdm} \sim 0.1$, although the
coincidence of closely related energy densities for baryons, CDM,
HDM and dark energy required would be amazing.  The $\omega_b$ and
$n_s$ likelihood curves are essentially independent of
$\Omega_{m\nu}/\Omega_m$ for fixed $\Omega_m$. \normalsize }
\label{fig:mnu}
%  \end{center}
\end{figure}

We use another 2 parameters to characterize the transport of the
radiation through the era of photon decoupling, which is sensitive
to the physical density of the various species of particles
present then, $\omega_j \equiv \Omega_j {\rm h}^2$. We really need
4: $\omega_b$ for the baryons, $\omega_{cdm}$ for the cold dark
matter, $\omega_{hdm}$ for the hot dark matter (massive but light
neutrinos), and $\omega_{er}$ for the relativistic particles
present at that time (photons, very light neutrinos, and possibly
weakly interacting products of late time particle decays). For
simplicity, though, we restrict ourselves to the conventional 3
species of  relativistic neutrinos plus photons, with
$\omega_{er}$ therefore fixed by the CMB temperature and the
relationship between the neutrino and photon temperatures
determined by the extra photon entropy accompanying $e^+ e^- $
annihilation. Rather than be exhaustive when the data does not
warrant it, the effect of massive neutrinos is shown in
Fig.~\ref{fig:mnu} for a few values of $ \omega_{hdm}/\omega_m$,
where $\omega_m \equiv \omega_{hdm}+\omega_{cdm}+\omega_b$.

 The angular diameter distance relation adds dependence upon
$(1-\Omega_{tot})$ and $\Omega_\Lambda$ as well as on $\Omega_m
\equiv  \omega_m / {\rm h}^2$ \cite{wQ}.  This defines a
functional relationship between $\Omega_{tot}$ and
$\Omega_\Lambda$, a  {\it degeneracy} \cite{degeneracies} that
would be exact except for the integrated Sachs-Wolfe effect,
associated with the change of $\Phi$ with time if $\Omega_\Lambda$
or $\Omega_k$ is nonzero. (If $\dot{\Phi}$ vanishes, the energy of
photons coming into potential wells is the same as that coming
out, and there is no net impact of the rippled light cone upon the
observed $\Delta T$.)

 Our 7th parameter is an astrophysical one, the Compton "optical depth" $\tau_C$ from a
reionization redshift $z_{reh}$ to the present. It lowers ${\cal
C}_\ell$ by $\exp(-2\tau_C)$ at the high $\ell$'s probed by
Boomerang. For typical models of hierarchical structure formation,
we expect $\tau_C \lta 0.2$. It is partly degenerate with
$\sigma_8$ and cannot be determined at this precision by CMB data
now.

The LSS also depends upon our parameter set: the most important
combination is the wavenumber of the horizon when the energy
density in relativistic particles equals the energy density in
nonrelativistic particles: $k_{Heq}^{-1} \approx 5 \Gamma^{-1}
\hmpc$, where $\Gamma \approx \Omega_m {\rm h}
\Omega_{er}^{-1/2}$. Instead of ${\cal P}_\Phi (k_n)$ for the
amplitude parameter, we often use ${\cal C}_{10}$ at $\ell =10$
for CMB only, and $\sigma_8^2$ when LSS is added; $\sigma_8^2$ is
a bandpower for density fluctuations at $8\hmpc$, a scale
associated with rare clusters of galaxies. When LSS is considered
in this paper, it refers to constraints on $\Gamma + (n_s-1)/2$
and $\ln \sigma_8^2$ that are obtained by comparison with the data
on galaxy clustering and cluster abundances \cite{lange00}.

When we allow for freedom in $\omega_{er}$, the abundance of
primordial helium, tilts of tilts ($dn_{\{s,is,t\}}(k_n)/d\ln k,
...$) for 3 types of perturbations, the parameter count would be
17, and many more if we open up full theoretical freedom in
spectral shapes. However, as we shall see, as of now only 3 or 4
combinations can be determined with 10\% accuracy with the CMB.

\begin{table*}
%[htb]
\caption{\small Cosmological parameter values and their 1-sigma
errors are shown, determined after marginalizing over the other
$6$ cosmological and $4^{+}$ experimental parameters, for
B98+Maxima-I+prior-CMB and the weak prior used in
\cite{lange00,jaffe00} ($0.45 \le {\rm h} \le 0.9$, age $> 10$
Gyr). The LSS prior was also designed to be weak. The detections
are clearly very stable if extra "prior" probabilities for LSS and
SN1 are included. (Indeed, they are stable to inclusion of
stronger priors -except if the BBN-derived $0.019 \pm 0.002$ is
imposed \cite{lange00}.) Similar tables for B98+DMR are given in
\cite{lange00} and for B98+MAXIMA-I+DMR in \cite{jaffe00}. If
$\Omega_{tot}$ is varied, parameters derived from our basic 7 come
out to be: age=$13.2\pm 1.3$ Gyr, ${\rm h}=0.70 \pm 0.09$,
$\Omega_m=0.35\pm .06$, $\Omega_b=0.065 \pm .02$. Restriction to
$\Omega_{tot}=1$ yields: age=$11.6\pm 0.4$ Gyr, ${\rm h}=0.80\pm
.04$, $\Omega_m=0.31\pm .03$, $\Omega_b=0.046 \pm .005$.
\normalsize } \label{tab:exptparams}
\hspace{-0.15truecm}
\begin{tabular}{|l|llll|llll|}
\hline
 {}  & cmb & +LSS & +SN1 & \small +SN1,LSS \normalsize & cmb & +LSS & +SN1 & \small +SN1,LSS \normalsize\\
\hline $\Omega_{tot}$            & $1.09^{+.07}_{-.07}$ &
$1.08^{+.06}_{-.06}$ & $1.04^{+.06}_{-.05}$ & $1.04^{+.05}_{-.04}$ & 1.0 & 1.0 & 1.0 & 1.0 \\
$\Omega_b{\rm h}^2$             & $.031^{+.005}_{-.005}$ &
$.031^{+.005}_{-.005}$ & $.031^{+.005}_{-.005}$ &
$.031^{+.005}_{-.005}$ & $.030^{+.004}_{-.004}$ &
$.030^{+.003}_{-.004}$
& $.030^{+.004}_{-.004}$ & $.030^{+.003}_{-.004}$ \\
$\Omega_{cdm}{\rm h}^2$ & $.17^{+.06}_{-.05}$ &
$.14^{+.03}_{-.02}$ & $.13^{+.05}_{-.05}$ & $.15^{+.03}_{-.02}$ &
$.19^{+.06}_{-.05}$ & $.17^{+.02}_{-.02}$
& $.16^{+.03}_{-.03}$ & $.17^{+.01}_{-.02}$ \\
$n_s$            & $1.05^{+.09}_{-.08}$ & $1.04^{+.09}_{-.08}$ &
$1.05^{+.10}_{-.09}$ & $1.06^{+.08}_{-.08}$ & $1.02^{+.08}_{-.07}$
& $1.03^{+.08}_{-.07}$ &
$1.03^{+.08}_{-.07}$ & $1.04^{+.07}_{-.07}$ \\
$\Omega_{\Lambda}$ & $0.48^{+.20}_{-.26}$ & $0.63^{+.08}_{-.09}$ &
$0.72^{+.07}_{-.07}$ & $0.70^{+.04}_{-.05}$ & $0.58^{+.17}_{-.27}$
& $0.66^{+.04}_{-.06}$ & $0.71^{+.06}_{-.07}$ & $0.69^{+.03}_{-.05}$ \\
\hline
\end{tabular}\\[2pt]
\end{table*}

Table~\ref{tab:exptparams} shows there are strong detections with
only the CMB data for $\Omega_{tot}$, $\omega_b$ and $n_s$ in the
minimal inflation-based 7 parameter set. The ranges quoted are
Bayesian 50\% values and the errors are 1-sigma, obtained after
projecting (marginalizing) over all other parameters. With Maxima,
$\omega_{cdm}$ begins to localize, but much more so when LSS
information is added. Indeed, even with just the COBE-DMR+LSS
data, the 2-sigma contours shown in Fig.~\ref{fig:dmde} are
already localized in  $\omega_{cdm}$. That $\Omega_\Lambda$ is not
well determined is a manifestation of the
$\Omega_{tot}$--$\Omega_\Lambda$ near-degeneracy discussed above,
which is broken when LSS is added because the CMB-normalized
$\sigma_8$ is quite different for open cf. $\Lambda$ models.
Supernova at high redshift give complementary information to the
CMB, but with CMB+LSS (and the inflation-based paradigm) we do not
need it: the CMB+SN1 and CMB+LSS numbers are quite compatible. In
our space, the Hubble parameter, ${\rm h}= (\sum_j (\Omega_j{\rm
h}^2 ))^{1/2}$, and the age of the Universe, $t_0$, are derived
functions of the $\Omega_j{\rm h}^2$: representative values are
given in the Table caption.

Since this is a conference devoted to neutrinos,
Fig.~\ref{fig:mnu} is included to show what happens as we let the
fraction of the matter in massive neutrinos vary, from 0 to 0.3.
Until Planck precision, the CMB data by itself will not be able to
strongly discriminate this ratio. Adding HDM does have a strong
impact on the CMB-normalized $\sigma_8$ and the shape of the
density power spectrum (effective $\Gamma$ parameter), both of
which mean that adding some HDM to CDM is strongly preferred in
the absence of $\Omega_\Lambda$. However a nonzero (though
smaller) $\Omega_\Lambda$ is still preferred.

We can also forecast dramatically improved precision with further
analysis of Boomerang and Maxima, other future LDBs, MAP and
Planck. Because there are correlations among the physical
variables we wish to determine, including a number of
near-degeneracies beyond that for $\Omega_{tot}$--$\Omega_\Lambda$
\cite{degeneracies}, it is useful to disentangle them, by making
combinations which diagonalize the error correlation matrix,
"parameter eigenmodes" \cite{bh95,degeneracies}.  For this
exercise, we will add $\omega_{hdm}$ and $n_t$ to our parameter
mix, making 9. (The ratio ${\cal P}_{GW}(k_n)/{\cal P}_\Phi (k_n)$
is treated as fixed by $n_t$, a reasonably accurate inflation
theory result.) The forecast for Boomerang based on the 440 sq.
deg. patch with a single 150 GHz bolometer used in the published
data is 3 out of 9 linear combinations should be determined to
$\pm 0.1$ accuracy. This is indeed what we get in the full
analysis of the Table and Boomerang+DMR only. If 4 of the 6 150
GHz channels are used and the region is doubled in size, we
predict 4/9 could be determined to $\pm 0.1$ accuracy. And if the
optimistic case for all the proposed LDBs is assumed, 6/9
parameter combinations could be determined to $\pm 0.1$ accuracy,
2/9 to $\pm 0.01$ accuracy. The situation improves for the
satellite experiments: for MAP, we forecast  6/9 combos to $\pm
0.1$ accuracy, 3/9 to $\pm 0.01$ accuracy; for Planck,  7/9 to
$\pm 0.1$ accuracy, 5/9 to $\pm 0.01$ accuracy. While we can
expect systematic errors to loom as the real arbiter of accuracy,
the clear forecast is for a very rosy decade of high precision CMB
cosmology that we are now fully into.

\normalsize

\end{document}